\documentclass[useAMS,usenatbib]{mn2e}
\usepackage{graphicx}
\usepackage{amsmath}
\usepackage{amssymb}
\usepackage{color}
\usepackage{url}

\newcommand  \acc     {\ifmmode {\rm km\,s}^{-2} \else km\,s$^{-2}$\fi}

\newcommand  \ergs     {\ifmmode {\rm ergs\,s}^{-1} \else ergs s$^{-1}$\fi}
\newcommand  \ergcms   {\ifmmode {\rm erg~cm}^{-2}\,{\rm s}^{-1}
                        \else erg~cm$^{-2}$\,s$^{-1}$\fi}
\newcommand  \ergcmsA  {\ifmmode{\rm erg\,cm}^{-2}\,{\rm s}^{-1}\,{\rm\AA}^{-1}
                        \else ergs\,cm$^{-2}$\,s$^{-1}$\,\AA$^{-1}$\fi}
\newcommand  \ergcmsHz {\ifmmode{\rm ergs\,cm}^{-2}\,{\rm s}^{-1}\,{\rm Hz}^{-1}
                        \else ergs\,cm$^{-2}$\,s$^{-1}$\,Hz$^{-1}$\fi}
\newcommand  \phcms    {\ifmmode {\rm ph\,cm}^{-2}\,{\rm s}^{-1}
                        \else ph\,cm$^{-2}$\,s$^{-1}$\fi}
\newcommand  \phcmsA   {\ifmmode {\rm ph\,cm}^{-2}\,{\rm s}^{-1}\,{\rm\AA}^{-1}
                        \else ph\,cm$^{-2}$\,s$^{-1}$\,\AA$^{-1}$\fi}

\newcommand\aj{{AJ}}%
\newcommand\apj{{ApJ}}%
\newcommand\apjl{{ApJ}}%
\newcommand\aap{{A\&A}}%
\newcommand\mnras{{MNRAS}}%
\newcommand\nat{{Nature}}%
\title[Searching for Galactic FRBs with radio receivers]
{Searching for giga-Jansky fast radio bursts from the Milky Way 
with a global array of low-cost radio receivers
}
\author[D. Maoz and A. Loeb]
{Dan Maoz$^{1}$, Abraham Loeb$^{1,2}$\\
{\small{\it $^{1}$~School of Physics and Astronomy, Tel-Aviv University, Tel-Aviv 69978,
Israel}}\cr
{\small {\it $^{2}$Institute for Theory and Computation, 
Harvard University, Cambridge, MA 02138, USA}}\cr
}\date{\today}

\begin{document}

\maketitle

\label{firstpage}

\begin{abstract}
If fast radio bursts (FRBs) originate from galaxies at cosmological
distances, then their all-sky rate implies that the Milky Way may host an
FRB every 30--1500 years, on average. 
If many FRBs persistently repeat for decades or more, 
a local giant FRB could be active now, with 
 1~GHz flux radio pulses
of $\sim 3\times 10^{10}$~Jy, comparable to the fluxes and
frequencies detectable by cellular communication devices (cell phones,
Wi-Fi, GPS).  
 We propose searching for Galactic FRBs using a global array
of low-cost radio receivers. One possibility is the $\sim
1$~GHz communication channel in cellular phones, through a
Citizens-Science downloadable application. Participating phones would
continuously listen for and record candidate FRBs
 and would periodically upload information to a
central data-processing website which will identify the signature of a
real, globe-encompassing, FRB from an astronomical distance.
Triangulation of the GPS-based pulse arrival times reported from different
Earth locations will provide the FRB sky position, potentially to
arcsecond accuracy. Pulse arrival times versus frequency, 
 from reports from phones operating at
diverse frequencies, or from fast signal de-dispersion by the
application, will yield the dispersion measure (DM). 
Compared to a Galactic DM model, it will indicate the source distance
within the Galaxy. A variant approach uses the
built-in $\sim 100$~MHz FM-radio receivers present in cell phones for
an FRB search at lower frequencies. Alternatively,
numerous ``software-defined radio'' (SDR) devices, costing $\sim$\$10
US each, could be deployed and plugged into USB ports of personal
computers (particularly in radio-quiet locations) to
establish the global network of receivers.

\end{abstract}


\newpage

\section{Introduction}
The origin and nature of fast radio bursts (FRBs) have remained
enigmatic since the first FRB discovery by \citet{lorimer07}.  The $
17$ or so distinct FRB sources that have been reported so far are
bright ($\sim 0.1$--$1~{\rm Jy}$) and brief ($\sim 1~{\rm ms}$) pulses
of $\sim 1~{\rm GHz}$ radio emission
\citep{lorimer07,keane12,thornton13,spitler14,burke-spolaor14,petroff15a,ravi15,champion16,masui15,keane16}.
The pulse arrival times of FRBs show a $\nu^{-2}$ frequency dependence
indicative of a passage through a cold plasma, with the so-called
dispersion measure (DM) measuring the line-of-sight column density of
free electrons.  FRBs are selected to have large measured DMs of $\sim
300-1600$~pc~cm$^{-3}$, in excess of the values expected from models
of the interstellar electron distribution in the Milky Way galaxy, and
have therefore been inferred to originate from extragalactic sources
at cosmological distances. The cosmological distance has been
confirmed in the case of the sole repeating FRB~121102, which has been
localised to a dwarf galaxy at redshift $z=0.19$ \citep{chatterjee,tendulkar,marcote}.

The range of excess (above Galactic) DMs of the known FRBs correspond
to a range of co-moving distances of 0.9--4.4~Gpc, with a median of
2.4~Gpc (corresponding to a redshift of $z=0.64$), under the
assumption that most of the excess DM is contributed by the
intergalactic medium and using the standard cosmological parameters
\citep{planck}.  The latest estimate of the all-sky rate of FRBs with
flux $>0.3$~Jy is $1.1^{+3.8}_{-1.0}\times 10^4~{\rm day}^{-1}$ at
95\% confidence \citep{scholz16}. This estimate is based on the single
detection of FRB~121102, and ignoring the fact that it has been
detected repeatedly over a period of 4 years. We note that some other
recent rate estimates have smaller uncertainties, but find, to varying
 degrees of significance, a dependence of rate on Galactic latitude
 (see, e.g., \cite{vanderwiel16}, and references therein). By adopting
 the \citet{scholz16} rate and its uncertainty, we encompass this
 uncertainty in the latitude dependence as well.  

If FRBs occur in
Milky-Way-like galaxies, we can divide the observed cosmological rate
by the number of galaxies within the FRB survey volume, to find the
expected FRB rate within a single such galaxy. We note that the
recently localised FRB~121102
comes from an extreme-emission-line dwarf galaxy \citep{tendulkar},
quite unlike the Milky Way. However, this host galaxy is not necessarily
representative of all FRB hosts.
The product
between the comoving number density of $L_*$ galaxies, $\sim
10^{-2}~{\rm Mpc^{-3}}$ \citep{prada}, and the cosmological volume out
to the median distance of known FRBs, $57$~comoving Gpc$^3$, implies
that an FRB should occur in our galaxy once per $140^{+1400}_{-110}$
years \citep{lingam17}.  A 0.3~Jy FRB from a comoving distance of $2.4$~Gpc 
(a luminosity distance of $3.9$~Gpc), placed at a typical Galactic
distance of $\sim 10$~kpc, would have an observed 1~GHz flux density
of $f_\nu\approx 3\times 10^{10}$~Jy or, equivalently, $3\times
10^{-16}~{\rm W~m}^{-2} {\rm Hz}^{-1}$. FRB~121102 has been bursting 
repeatedly for at least 4 years. If most FRBs persist for decades or 
even centuries, a Galactic FRB could be active now. A powerful
local FRB may have already been detected in the far side-lobes of radio
telescope beams, but mistakenly ascribed to artificial interference.
 
Indeed, the radio flux density level of the received GHz-band signals from
commercial radio stations, cellular communications and wireless
networks is within a few orders of magnitude of the expected flux
level from a Galactic FRB.  For example, a typical desktop Wi-Fi
transmitter operating at 2.4~GHz under the 802.11b standard has a
radiated power of 100~mW over an 82~MHz bandpass with an outdoor range
of $\sim 100$~m, corresponding to a detected flux density of
$f_\nu=1\times 10^{-14}~ {\rm W~m}^{-2} {\rm Hz}^{-1}$. Each of the
transmitter's individual channels has a bandpass of 22~MHz, and
therefore a time resolution of $\Delta t \sim 2\times 10^{-8}$~s. By binning an
incoming signal into millisecond ($\Delta t=10^{-3}$~s) time bins
a Wi-Fi receiver would improve its sensitivity in
proportion to $\sqrt{\Delta t}$, i.e. by a factor of $\sim
200$, to a level of $f_\nu\sim 5\times 10^{-17}~ {\rm W~m}^{-2} {\rm
  Hz}^{-1}$ ($5\times 10^9$~Jy, i.e. 5~GJy). This is a factor 6
 fainter than the typical Galactic FRB flux discussed above,
and means that such a Galactic FRB, and even fainter and
perhaps-more-frequent FRBs, would be detectable by existing
communication devices. In the subsequent sections, we outline how an
array of numerous low-cost radio receivers can be used to detect and
localise giant Galactic FRBs.
  
\section{A global array of cellular receivers for Galactic FRB detection}

We consider below three related technical approaches to the
assembly of an array of low-cost radio receivers, suitable for the
detection of Galactic FRBs. The choice of the most practical approach
will depend on several issues that need to be resolved, such as the
ability to access and manipulate raw radio signals picked up by the
antennas, the flux from FRBs at sub-GHz frequencies, the level of
terrestrial noise at different locations, and the ability to filter
out that foreground noise.

\subsection{A cell phone communications channel approach}

There are currently an estimated 7 billion active cellular phone
accounts on our planet (similar to the number of people), operating in
several frequency bands, from 0.8 to 2.4~GHz. Each of these phones is,
as argued above, a radio receiver that is in principle sensitive to a
Galactic FRB signal. Furthermore, every smartphone is a programmable
computer capable of analyzing the signal, of timing it up to $\Delta t
\sim 10^{-7}$~s precision with its global-positioning system (GPS)
module, of storing this information, and of diffusing it through the
internet.

We propose therefore to build a Citizens-Science project in which
participants voluntarily download onto their phones an application
that runs in the background some or all of the time, monitoring the
phone's antenna input for candidate broad-band millisecond-timescale
pulses that appear similar to an FRB.  The application would record
candidate FRB pulses (most of which originate from artificial and
natural noise sources) and would periodically upload the candidate
pulse information (pulse profile, GPS-based arrival time), along with
information about the phone (GPS-based location, operating frequency)
to a central processing website. The central website will continuously
correlate the incoming information from all participants, to identify
the signature of a real, globe-encompassing, FRB.

Because of the received signal's integration into ms time bins
(required to improve the sensitivity to FRB levels, see above), every
phone's actual arrival time accuracy will be no better than
$10^{-3}$~s. However, improved time precision can be recovered by
averaging the reported arrival times recorded by many participating
phones at a similar location. For example, averaging the ms-precision
reports from 10,000 phones within a city of radius 3~km (light travel
time $<10^{-5}$~s), would improve the precision by a factor of 100, to
$\sim 10^{-5}$~s.  Triangulation of the GPS-timed pulse arrival times
from different Earth locations would then give the FRB sky position to
an accuracy of order $\sim c\Delta t/2R_\oplus\sim 1$~arcmin. If time
binning of the FRB signal, and subsequent loss of the native
$10^{-7}$~s GPS timing precision could be avoided (a possibility
considered in some of the other technical frameworks that we propose
below), then naturally the localisation precision can be improved down
to
the sub-arcsecond level.

Because of the $\nu^{-2}$ arrival-time dependence of a radio pulse
propagating through the Galactic plasma, phones operating at diverse
frequencies (multiple networks and phone models) will receive the
signal at a time delay,
\begin{equation}
\delta t= 0.144\times\left({{\rm DM}\over 200~{\rm
    pc~cm^{-3}}}\right)\left({\nu\over {\rm 2.4 GHz}}\right)^{-2} ~{\rm s} .
\end{equation} 
Over, e.g., a 22~MHz cellphone channel bandwidth at 2.4~GHz, a typical
Galactic DM of $200~{\rm pc~cm^{-3}}$ \citep{Rane_16} will spread  the 
FRB arrival time over just 2.6~ms, comparable to typical FRB pulse widths. 
The channel bandwidth 
therefore will not result in any significant smearing of 
the pulse over time, which could have reduced the
detection sensitivity and timing precision.
By comparing the arrival times of different frequencies at the same
locations, the central website will be able to solve for the FRB's DM
that, when compared to a Galactic DM model \citep{cordes02,yao16}, will
indicate the FRB source distance within the Galaxy. Alternatively, the 
application software itself could attempt to de-disperse all candidate
incoming signals across the full frequency range available to each
receiver. With efficient new algorithms, real-time de-dispersion of FRB signals is now feasible
on small computers \citep{Zackay_14,Zackay_17}, and so is likely
possible on smartphones as well. In such a scenario, the
identification of a $\nu^{-2}$
frequency sweep would be a real-time test of incoming signals, performed at the 
level of each individual receiver.
 
One clear advantage of the above operation plan is that it is
essentially cost free---all of the necessary hardware (the world's
cell phones) is already in place, and one needs only to carry out the plan's
organizational steps in order to make it work for the scientific
program.  Potential problems with this proposed mode are, first, that
cell phones may be hardwired at the basic electronics level to
demodulate and digitize incoming communications signals, and therefore
the raw broad-band radio signal containing the FRB may be inaccessible
to software. Furthermore, mobile phone communications are encoded so
as to allow many users to share the frequency band, and this encoding 
permits the detection of communication signals at sub-noise levels 
(as opposed to the un-encoded FRB signal). 

The sought-after millisecond-timescale FRB
signal will need to be disentangled from 
the foreground noise of cellular and other
communications emissions, as  well as from natural radio noise from
atmospheric processes and from the sun. Although the feasibility of
this requires further study, the prospects look promising based on 
a number of past attempts. \citet{katz2003} 
review a handful of experiments, and decribe their own experiment, 
which is similar to our proposal. 
The basic concept consisted of a number of wide-angle, geographically distributed
radio receivers that searched for short radio bursts, separating
astronomical signals from noise by requiring coincident detections. 
\citet{katz2003} used three 611~MHz receivers in the eastern US, sensitive to
$\gtrsim 3\times 10^4$~Jy bursts ($2\times 10^5$ fainter than considered here)
on timescales $\gtrsim 125$~ms (50 times longer than here). 
The recorded, GPS-time-stamped, bursts were periodically uploaded to a
central processing station, exactly as in our proposed plan.

Over 18 months of continuous operation, \citet{katz2003}
 detected a burst roughly
every 10~s, but 99.9\% of these signals could be rejected as local
interference based on their non-coincidence between the three receivers. 
The remaining $\sim 4000$ coincident signals could all be traced to
solar radio bursts, by comparison to reports from a solar radio observatory.
No other astronomical sources were detected by \citet{katz2003} nor by
previous experiments. Interestingly, \citet{katz2003} succeeded in 
using their GPS signal, with its $10^{-7}$~s accuracy, to time-stamp 
their detected bursts to the accuracy of their 
20~$\mu$s-long individual time samples, and they note
that, in principle, they could have used a multiple-time-sample 
averaging period
shorter than 0.125~s (at the expense of sensitivity). 
This would have allowed them to triangulate
their source localisations, just as we propose to do.  

\subsection{A cell phone FM radio  channel approach}  

Most or all cell phones have built-in FM-band radio receivers
operating at around $\nu\sim 100$~MHz and enabling direct (i.e. not
through the internet or the service provider) reception of radio
broadcasts. Interestingly, this hardware is de-activated by phone
manufacturers in about $\sim {2\over 3}$ of all phones, in the
service-providers' interest of having the customers download and pay
for the radio broadcasts, rather than receiving them for free.
Nevertheless, about ${1\over 3}$ of all phones (still a sizeable
number when considering the global number) do have the direct FM
reception option activated. The raw, non-demodulated radio signal from
this channel is more likely to be accessible to the application
software in its search for an FRB signal than in the preceeding 
approach using the $\sim 1$~GHz cellular communication channel. A shortcoming of this
option, however, is the yet-unknown properties of FRBs at $\sim
100$~MHz frequencies. Current upper limits from FRB searches at
145~MHz \citep{karastergiou} and 139-170~MHz \citep{tingay}, limit the
FRB spectral slope to $>+0.1$. 
  As with the $\sim 1$~GHz cellular-communications option,
discussed above, here too integration over time could
make a typical Galactic FRB 
detectable at 100~MHz,
even for more positive slopes as high as $+1$, such that
the FRB would have $\gtrsim 5$~GJy at 100~MHz.  The foreground noise
question in this option is similar (though in a different frequency
band) to that in the previous, cellular-communications, option.

\subsection{A software-defined radio approach}   

A software-defined radio (SDR) is a radio system where components such
as filters, amplifiers, demodulators, etc., that are typically
implemented in hardware, are implemented instead in software on a
personal computer. SDR devices are widely available for $\sim$\$10~US
a piece, and they are popular with radio amateurs. 
They are often the size of a memory stick and likewise can be
USB plugged. An SDR device includes an antenna than can detect the
full raw ambient radio emissions over some frequency range and can
input them with minimal processing into a computer, where the signals
can be software-processed at will. Our third approach is therefore to
deploy a large number (depending on the available budget) of such SDR
devices, to be plugged into participating personal computers around
the globe, or base the network on devices already in use by
participating radio amateurs. As with the phone option, the participants will download
and install software that will continuously monitor the input from the
SDR. As before, the computers will upload the information on candidate
Milky Way FRBs to a central data-processing website.

A disadvantage of this approach is the need to actually buy and send
the SDR hardware to the selected participating individuals of the
network (unless one takes the existing-amateur-SDR approach). 
The advantages involve having an accessible FRB signal,
uniformly processed and fully analysable at will (including spectral
information from every station). Every SDR could be supplemented with
a simple exterior antenna or antenna booster (wireless reception
boosters are also widely and inexpensively available for cell phones
and laptops) that would considerably enhance its sensitivity, lowering
or fully avoiding the need for time integration, and
hence for the sacrifice of timing precision, or simply probing 
for fainter and more frequent bursts (see below).   Furthermore, the ability
to choose the stations sites at will in a well-spaced global network,
specifically in ``radio-quiet'' locations with minimal artificial and
natural radio interference, may prove to be the most important benefit.

\section{Lower-flux, more-common Galactic FRBs}

A major practical problem of the schemes described above are the long
and uncertain timescales---decades to many centuries---expected for
the detection of a single, Galactic $3\times 10^{10}$~Jy FRB, unless
typical FRBs persistently repeat for decades or centuries (which is
a real possibility, given the case of FRB~121102). If FRBs typically
do not repeat, then even for
the more optimistic end of the rate estimate, broadcasting standards,
phone models and other technical factors, may change over a decade,
not to mention the limited patience of the participants and the
experiment managers. A resolution of this concern, however, could be
based on the fact that FRBs must have a distribution of
luminosities. Indeed, if the known FRBs are at the cosmological
distances indicated by their excess DMs, then they are clearly not
``standard candles''. A reasonable expectation is then that FRB
numbers increase at decreasing luminosities. If so, lower-luminosity
FRBs should be detected more frequently by the global cellular
network.

Let us assume that we can parameterize the FRB number per unit 
luminosity with a Schechter form,
\begin{equation}
{dN \over d(\log L_\nu)} \propto L_\nu^{-\alpha+1}e^{-L_\nu/L_{F\star}}, 
\end{equation}
where $L_{F\star}$ corresponds to the characteristic specific luminosity
of an FRB source (namely, the one that yields an observed flux density
of $\sim 0.3~$Jy at a luminosity distance of $\sim 4$~Gpc).  One way
to calibrate $\alpha$ is by speculating that the Galactic population
of rotating radio transients
(RRATs), which have some properties in common with
FRBs, constitute the low-luminosity counterparts of
FRBs. The rate of RRATs over the entire sky  at a flux
of $\sim 0.3$~Jy is $\sim 10^6~{\rm
  day}^{-1}$, based on the estimated number of sources in the Galaxy, 
$\sim 10^5$, and their individual repetition rates, 
$\sim 10$~day$^{-1}$\citep{mclaughlin06}.
 The RRAT rate is thus $\sim
10^{11}$ times the Galactic FRB rate (of once per 300~yr, i.e. 
$10^{-5}$~day$^{-1}$). The RRAT flux
of $\sim 0.3$~Jy, in turn, corresponds to $\sim 10^{-11}$ the typical flux of
a Galactic FRB. If these two populations of
transient radio sources are related, then  $\alpha\approx 2$. 
Interestingly, this value
corresponds to an equal luminosity contribution from transients per
logarithmic interval of luminosity.

The flux distribution from a Galactic
 FRB population having a particular  luminosity 
will be $(dN/d(\log f_\nu))\vert_{L_\nu}
\propto f_\nu^{-3/2}$ for a spherically distributed population 
(e.g. in the Galactic halo), or $\propto f_\nu^{-1}$ 
for a planar distribution (e.g. the Galactic disk)---
 coincidentally matching the power-law scaling at
low fluxes in the luminosity function for $\alpha=2$. 
At a 5~GJy flux level, still detectable by our proposed arrays,
one might then expect to find Galactic FRBs 6 times more frequently than 
at 30~GJy, i.e. once per 5 to 250 years. Increasing the sensitivity
by one or two orders of magnitudes, e.g. by adding simple antennas
in the SRD option,  would potentially  allow for the detection of 
Galactic FRBs on a yearly to weekly basis, and for the direct 
determination of their luminosity function.

\section{Conclusions}

The first, and so-far only, FRB that has been localised, FRB~121102,
is at a cosmological distance, it has been repeating for at least 4
years, and its host galaxy is a low-metallicity dwarf.
We have argued that if most FRBs are cosmological, but their hosts
are not necessarily dwarf galaxies like the host of FRB~121102, 
then their all-sky rate implies that the Milky Way hosts an FRB
every 30 to 1500 years. If, furthermore, many FRBs repeat like FRB~121102, and for long enough, then
the occurrence frequency could be higher, and a local FRB may even be
active now. A typical Galactic FRB will be a
millisecond broad-band radio pulse with 1~GHz flux density of $\sim
3\times 10^{10}$~Jy, not much different from the radio flux levels and
frequencies detectable by cellular communication devices (cell phones,
WiFi, GPS). If the Milky Way has a currently active and repeating FRB source, 
then some Local Group galaxies would have them too, at MJy
flux-density levels, which 
could be detected by monitoring nearby galaxies with dedicated
small radio telescopes.   

An argument against frequent Galactic FRBs could be that FRBs require
some kind of exotic and energetic physical event, such as a
super-flare from a magnetar, and that irradiation of the Earth 
by such an event once per
century or millenium would be accompanied by clear signatures, or
perhaps even by mass extinctions. However, this argument relies on a
still-speculative connection between the radio emission of FRBs and their
emissions in other bands. Observationally, an upper limit of $10^8$~Jy
has been set on any FRB-like radio flux accompanying the giant 2004 December
$\gamma$-ray burst from the magnetar SGR~1806-20, and no $\gamma$-ray 
counterparts have been detected for any FRB \citep{tendulkar16}.

Our proposed search for Galactic FRBs using a global array of low-cost
(possibly already existing) radio receivers would enable triangulation
of the GPS-timed pulse arrival times from different Earth locations,
localising the FRB sky position to arcminute or even arcsecond
precision. Pulse arrival times from devices operating at diverse
frequencies, or from de-dispersion calculations on the devices
themselves,
will yield the DM that, when compared to a Galactic DM
model, will indicate the FRB source distance within the Galaxy.
Fainter FRBs could potentially be detected on a yearly or even weekly 
basis, enabling a direct measurement of the FRB luminosity function.
   
\section*{Acknowledgments}
We thank C. Carlsson, A. Fialkov, J. Guillochon, Z. Manchester,
E. Ofek, M. Reid, B. Zackay, and the anonymous referee, for
useful advice and comments. This work was supported in
part by Grant 1829/12 of the I-CORE program of the PBC and the Israel
Science Foundation (D.M.)  and by a grant from the Breakthrough Prize
Foundation (A.L.).  A.L. acknowledges support from the Sackler
Professorship by Special Appointment at Tel Aviv University.

\bibliography{wdbib} \bibliographystyle{apj}

\end{document}